\documentclass{article}

\usepackage[resetlabels,labeled]{multibib}
\newcites{S}{Primary Sources Cites}

\usepackage{color}

\usepackage{xcolor,colortbl}
\xdefinecolor{gray95}{gray}{0.65}
\xdefinecolor{gray25}{gray}{0.8}

\usepackage{authblk}

\title{Personalization of Computer-Based Technologies for Autism: \\ An Open  Challenge for Software Engineering?}
\author[1]{Roberto E. Lopez-Herrejon}
\author[2]{Gerardo Herrera}
\author[2]{Javier Sevilla}
\affil[1]{Dept. Software Engineering and IT, \'{E}cole de technologie sup\'{e}rieure, University of Qu\'{e}bec, Canada. roberto.lopez@etsmtl.ca}
\affil[2]{University of Valencia, Spain. \{gerardo.herrera,javier.sevilla\}@uv.es}
\date{}                     
\begin{document}

\maketitle

\begin{abstract}
Autism Spectrum Disorder (ASD) is neurodevelopmental condition characterized by social interaction and communication difficulties, along with narrow and repetitive interests. Being an spectrum disorder, ASD affects individuals with a large range of combinations of challenges along dimensions such intelligence, social skills, or sensory processing. Hence, any computer-based technology for ASD ought to be personalized to meet the particular profile and needs of each person that uses it.
Within the realm of Software Engineering, there is an extensive body of research and practice on software customization whose ultimate goal is meeting the diverse needs of software stakeholders in an efficient and effective manner.
These two facts beg the question: Can computer-based technologies for autism benefit from this vast expertise in software customization? As a first step towards answering this question, we performed an exploratory study to evaluate current support for customization in this type of technologies. 
Our study revealed that, even though its critical importance, customization has not been addressed. We argue that this area is ripe for research and application of software customization approaches such as Software Product Lines.
\end{abstract}


\section{Introduction}
\label{sec:introduction}

Autism Spectrum Disorder (ASD) is neurodevelopmental condition characterized by social interaction and communication difficulties, along with narrow and repetitive interests. 
ASD affects individuals in multiple and combined ways along 
areas such intelligence, social skills (e.g. unable to interpret non-verbal cues), or sensorial processing (e.g. sensitivity to noise or lights)~\cite{ANYAS/ElKalioubyPB06}. In the autism community, a common saying is: \emph{"if you've met one person with autism you've met one person with autism"}\footnote{Quote authored by Stephen Shore, http://www.autismasperger.net/.}. This entails that individuals with autism have unique sets of challenges and needs that must be addressed to help their development and integration to society.
 
There is an extensive and long standing research on using computer-based systems, that spreads over more than four decades~\cite{Kientz2013}, whose driving goal is to support the needs of people with autism and their families.  Currently, digital libraries have hundreds of articles on the subject. This research has been summarized to certain degree in many literature reviews and surveys studies (e.g.~\cite{SAGE/GrynszpanWPG14,Stephenson2015}).
However, and despite its critical importance, there has not been a study on how this type of computer-based systems handle customization, also referred to in the autism community and literature as \textit{personalization} . 
In this paper, we address this issue by analyzing the customization capabilities of approaches published over the last five years. 
Furthermore, we want to collect information about the strength of the empirical evidence that supports each of the approaches. In other words, the types of formal research studies that have been performed with them.

Our exploratory study indeed corroborated the lack of research on customization for computer-based technologies for autism and the slim empirical evidence that supports them. 
Based on these findings, we argue that this area is ripe for research and application of advanced customization approaches such as Software Product Lines.
We conclude by sketching some first challenges in this area.


\nociteS{S2-PUCom/SitdhisanguanCDO12}
\nociteS{S3-PUCom/Keay-BrightW12}
\nociteS{S9-DISC/HailpernHKBK12}
\nociteS{S13-RosenbloomMWM16}
\nociteS{S14-MechlingAFB15}
\nociteS{S15-LiuSVS17} 
\nociteS{S16-TorradoGM17}
\nociteS{S17-CabiellesPPF17}
\nociteS{S18-ChevalierMIBT17}
\nociteS{S19-YingSA16}
\nociteS{S20-KhanPHHSM16}
\nociteS{S21-SharmaSAVHKTR16}
\nociteS{S22-MagriniSCC16}
\nociteS{S23-HulusicP16}
\nociteS{S24-SturmPP16}
\nociteS{S25-ChevalierIMT16}
\nociteS{S26-LandowskaS16}
\nociteS{S27-SkillenDNB16}
\nociteS{S28-KhoslaNC15}
\nociteS{S29-MeiMQ15}
\nociteS{S30-VoonBMJA15}
\nociteS{S31-QidwaiSHM14}
\nociteS{S32-GruarinWB13}
\nociteS{S33-DimitrovaVB12}

\section{Autism Background}
\label{sec:autism}

Autism Spectrum Disorder (ASD) is a neurodevelopmental disorder characterized by impaired social interaction and communication, and restricted and repetitive behavior~\cite{APA}. ASD is diagnosed in at least 1\% of the population, and diagnoses are more common amongst males than females~\cite{Baird2006}. Autism can have profound impact upon learning and it is estimated that 54\% of individuals with autism also have intellectual disability/learning disability (Center for Disease Control CDC\footnote{Center for Disease Control (CDC) https://www.cdc.gov/ncbddd/autism/data.html}). 

Today, no medical treatment is available for the core symptoms of autism. Early intervention programs, usually aimed at children from 0 to 6 years old,  have been demonstrated effective for supporting the development of a relevant percentage of children with autism. 
The most effective programs have a behavioural base (e.g. Applied Behaviour Analysis (ABA)~\cite{Cohen2006}) or cognitive-behavioural base (e.g. Early Start Denver Model (ESDM)~\cite{Rogers2010}). 
Within this later program, for example, there is a developmental curriculum (the ESDM checklist) that is highly personalised for each child with autism and intervention objectives are redefined every three months in order to adapt to the child progress. 

Intervention programs in autism can be classified as focused-intervention (FI) programs~\cite{Wong2015}, usually designed for improving a particular ability (or a reduced set of abilities) or comprehensive treatment models (CTM) that are much wider and are based on a holistic approach of the child development~\cite{Odom2010}. Some of these programs use technologies as a basis for documenting the child progress, and some other use technology for very particular tasks. However, none of these programs are genuinely based on any particular technology. When available, innovative technologies are used for Focused Interventions rather than as Comprehensive Treatment Models. Most research evidence available on technologies for ASD rely mainly on the use of particular communicator apps on tablet devices while the evidence on other areas seems to be anecdotal or at least not enough explored~\cite{Lorah2014}.



\section{Exploratory Study}
\label{sec:study}

The goal of our exploratory study is to gauge at the support of customization on computer-based technologies for autism. Hence, to retrieve the relevant literature we performed a search using Web of Science\footnote{http://apps.webofknowledge.com/}. We employed this search engine because its advanced query capabilities and because it indexes all the publication outlets on autism and technology. In this search, in addition to the term \textit{customization}, we also use the term \textit{personalization} as it is also commonly used in the autism literature. We constrained our query to sources published in the last five years. The query we employed was\footnote{TS=Topic Search, ASC=Autism Spectrum Condition}:
\begin{center}
	\begin{tabular}{p{10cm}}
		\texttt{
			(TS=((Autis* OR ASD OR ASC OR "Asperger Syndrome" OR "Pervasive Developmental Disorder" OR PDD*) AND (Technolog* OR Computer* OR
			Virtual* OR Robot*) AND (Custom* or Personali*))) AND (Search Language = English) AND (Document type= Article) AND (Timespan= 2012-2017) AND (Indexes: SCI-EXPANDED, SSCI, A\&HCI, CPCI-S, CPCI-SSH, ESCI, CCR-EXPANDED, IC.)
		}
	\end{tabular}
\end{center}

\noindent \textbf{Inclusion and exclusion}. The basic criterion for inclusion in our study was a clear application of a computer-based technology for supporting a therapy or intervention in relation to autism, where individuals with autism participated in the design, validation or evaluation of the technology. 
The criteria to exclude papers in our study was:
\textit{i)} papers which did not describe a technology that supports any intervention or therapy (e.g. paper that describes biosignal monitoring tool), 
\textit{ii)} individuals with autism were not involved at any stage of design, validation or evaluation of the technology, 
\textit{iii)}
papers not written in English, \textit{iv)} vision or position papers that had no implementation to back them up, \textit{v)} graduate or undergraduate dissertations and thesis, and \textit{vi)} non peer-reviewed documents such as technical reports.

During the screening process we looked for the search terms in the title, abstract and keywords and whenever necessary at the introduction or at other places of the paper. 
The decision on whether or not to include a paper was most of the times straightforward, in other words, a clear application of computer-based technologies to autism with the participation of individuals with the condition was easily drawn. 

The search query obtained 179 sources. After a careful sieving using the inclusion and exclusion criteria, our search produced 24 primary sources that we analyzed in further detail as presented in Table~\ref{tab:tech-int-comp}, which shows the type of technology used, the forms of interactions, and the support for customization. From this table, we can observe : \textit{i)} the pre-eminent technologies are mobile and smartphone devices, followed by robots; \textit{ii)} the predominant use of touch screen as form of interaction; and \textit{iii)} the dire lack of customization support, with 11 out of the 24 actually providing examples of customization and not simply mentioning it as desirable property. 


\begin{table*}[th!]
\center
\caption{Technology, interaction forms, and customization support summary}
\footnotesize
\begin{tabular}{p{1.0cm}|p{3.0cm}|p{3.0cm}|p{3.5cm}}
\hline
 \textbf{Primary source} & \textbf{Technology} & \textbf{Interaction forms} & \textbf{Customization support} \\ \hline 


 \citeS{S2-PUCom/SitdhisanguanCDO12} & tangible user interfaces &  object manipulation
 &  no support provided \\ \hline

 \citeS{S3-PUCom/Keay-BrightW12} & multi-touch screen & touch screen &  no support provided \\ \hline
  
 
 
 
 
 
 \citeS{S9-DISC/HailpernHKBK12} & speech recognition and visual feedback  &  monitors and microphones & no support provided \\ \hline
 
 
 

 \citeS{S13-RosenbloomMWM16} & smartphone &  screen interaction & customization for prompts, recording, data monitoring \\ \hline  
 
 \citeS{S14-MechlingAFB15} & video recording and playing &  video watching & highlights importance of custom-made videos  \\ \hline             

\citeS{S15-LiuSVS17} & smart glasses, augmented reality, games & eye gaze, movement & sensor data, game difficulty and rewards \\ \hline

\citeS{S16-TorradoGM17} & smartwatches, smartphones & app, smartwatch screen & authoring tool for coping strategies \\ \hline

\citeS{S17-CabiellesPPF17}  & tablets & tablet screen &  sequences, words \\ \hline

\citeS{S18-ChevalierMIBT17} & avatars, robots &  tacticle computer game &  no support provided \\ \hline

\citeS{S19-YingSA16} & mobile devices, smartphone & touch screen, mobile phone &  default avatars with personal pictures \\ \hline

\citeS{S20-KhanPHHSM16} & mobile devices & touch screen, mobile phone & no support provided \\ \hline

\citeS{S21-SharmaSAVHKTR16} & gesture detection & gestures & no support provided \\ \hline
\citeS{S22-MagriniSCC16} & gesture detection, body tracking & gestures, movement & sound selection for gestures and movement \\ \hline

\citeS{S23-HulusicP16} & web game & web page & not support provided \\ \hline
\citeS{S24-SturmPP16} & mobile devices & touch screen, mouse & therapy session contents \\ \hline
\citeS{S25-ChevalierIMT16} & robots & gestures, movement & 3 groups based on kinetic and propioception profiles \\ \hline

\citeS{S26-LandowskaS16} & tablets & touch screen & activities plan \\ \hline
\citeS{S27-SkillenDNB16} & smartphone & touch, visual & colors scheme, logging info, geo-fences \\ \hline
\citeS{S28-KhoslaNC15} & robots & speech, touch screen & activities and care services \\ \hline
\citeS{S29-MeiMQ15} & virtual reality, body tracking & motdio detection, touch screen, screen & avatars in game \\ \hline
\citeS{S30-VoonBMJA15} & mobile devices, tablet & touch screen, sound, visual & images, stepwise description of tasks \\ \hline

\citeS{S31-QidwaiSHM14} & robots & motion detection, tablet &  play activities \\ \hline
\citeS{S32-GruarinWB13} & sensors & tangible carpet & tasks and photographs \\ \hline
\citeS{S33-DimitrovaVB12} & robots & movement & no support provided \\ \hline


 \hline 
 \end{tabular}
\label{tab:tech-int-comp}
\end{table*}

\noindent \textbf{Empirical evidence support}. This refers to the empirical evidence that supports each of the sources identified, which we describe in terms of the research designs (or experimental designs) employed. 
We have considered the two principal types of design that are applied in most technological studies~\cite{Kientz2013,WohlinExperimentalSE12}. 
\textit{Single subject} research design refers to research in which the subject serves as his/her own control, rather than using another individual/group. 
\textit{Group research design} refers to research where one group of participants (treatment group) is compared to another group (control group) with participants in both groups balanced around variables such as age, IQ or severity of autism symptoms around social communication or restrictive/repetitive behaviours.
Additionally, we classify also the availability of the resources for replication, for instance open sources and documentation. For this latter category, we use \emph{none}, \emph{partial}, and \emph{full} depending on the degree of availability.

Table~\ref{tab:empirical} summarizes our findings. It is immediately clear that the majority of sources utilize the most basic type of design, single subject, with a very reduced number of participants, and only in one case all the information for replication is provided. Furthermore, five sources utilise group designs but the lack of available details hinder their replication. Finally, other five sources did not make any empirical evaluation or at least they did not describe the study in enough detail.  

\begin{table}[h]
\center
\caption{Empirical support summary}
\footnotesize
\begin{tabular}{p{4.0cm}cp{2.5cm}}
\hline

\cellcolor{gray25} Single subject design & \cellcolor{gray25} No & \cellcolor{gray25} Level \\ \hline

\citeS{S27-SkillenDNB16} &  16 & None \\
\citeS{S3-PUCom/Keay-BrightW12}	&  13 &  Partial \\
\citeS{S17-CabiellesPPF17} & 11 & Partial \\
\citeS{S29-MeiMQ15} & 10 &  Partial \\  
\citeS{S19-YingSA16} & 5 & Partial \\
\citeS{S14-MechlingAFB15}, \citeS{S22-MagriniSCC16}	& 4	& Partial \\

\citeS{S26-LandowskaS16} & 2 & Full \\

\citeS{S9-DISC/HailpernHKBK12}, 
\citeS{S15-LiuSVS17}, \citeS{S16-TorradoGM17}, \citeS{S17-CabiellesPPF17}, \citeS{S20-KhanPHHSM16} & 2 & Partial \\

\citeS{S28-KhoslaNC15} & 2 & None \\

\citeS{S32-GruarinWB13}	& 1	& Partial \\

\citeS{S13-RosenbloomMWM16}	& 1	& None \\

\hline 
\cellcolor{gray25} Group design & \cellcolor{gray25} No & \cellcolor{gray25} Level \\ \hline
\citeS{S2-PUCom/SitdhisanguanCDO12} &	32	& None \\

\citeS{S33-DimitrovaVB12} & 22 & None \\


\citeS{S18-ChevalierMIBT17}, \citeS{S25-ChevalierIMT16} & 13 & Partial \\
\citeS{S21-SharmaSAVHKTR16} & 10 & Partial \\ 

\hline 
\cellcolor{gray25} Not possible to determine & \cellcolor{gray25} No & \cellcolor{gray25} Level \\ \hline
\citeS{S1-IWVR/ParesCDFFGS04} & 11 & Partial \\ 

\citeS{S31-QidwaiSHM14}  & NA & Partial \\

\citeS{S23-HulusicP16}, \citeS{S24-SturmPP16}, \citeS{S30-VoonBMJA15} & NA & None \\

 \hline 
 \end{tabular}
\label{tab:empirical}
\end{table}

\section{Challenges for Software Engineering}

One of the leading approaches for software customization are \textit{Software Product Lines (SPLs)} which are families of related systems whose members are distinguished by the set of features they provide, where a \textit{feature} is an increment in functionality~\cite{DBLP:journals/tse/BatorySR04,SPLE}.
A key concept in SPLs is \textit{variability} which is the capacity of software artifacts to vary.  
Several forms of \textit{variability models} have been proposed that succinctly and formally express all the desired combination of features that the products of an SPL may have. Our study has revealed the following two open challenges:
\begin{itemize}
	\item Develop user profile models that formally describe all the variations that persons with autism may have along dimensions like sensory needs, intellectual disability, etc. taking information from the standard battery of tests used to diagnose the condition.
	
	\item Provide tool support to collect and analyze data that integrates with the standard workflow developers use (e.g. a plug-in for Android studio when developing apps) for improving the empirical evaluation of this type of technologies.
	
\end{itemize}


\section{Acknowledgments}
This study has been posible thanks to the funding received under the grant agreement 2015-1-ES01-KA201-015946 of the Erasmus+ Program of the European Union and the project Recherche interdisciplinaire sur les syst\`{e}mes logiciels variables sponsored by the \'{E}cole de technologie sup\'{e}rieure, University of Qu\'{e}bec, Canada.

%
\bibliographystyle{abbrv}
\bibliography{biblio}  

\begin{thebibliography}{10}

\bibitem{S2-PUCom/SitdhisanguanCDO12}
K~Sitdhisanguan, N.~Chotikakamthorn, A.~Dechaboon, and P.~Out.
\newblock {Using tangible user interfaces in computer-based training systems
  for low-functioning autistic children}.
\newblock {\em Personal and Ubiquitous Computing}, 16(2):269--285, 2012.

\bibitem{S3-PUCom/Keay-BrightW12}
W~Keay-Bright and I~Howarth.
\newblock {Is simplicity the key to engagement for children on the autism
  spectrum?}
\newblock {\em Personal and Ubiquitous Computing}, 16(2):209--221, 2012.

\bibitem{S9-DISC/HailpernHKBK12}
J~Hailpern, A~Harris, R~{La Botz}, B.~Birman, and K.~Karahalios.
\newblock {Designing visualizations to facilitate multisyllabic speech with
  children with autism and speech delays}.
\newblock In {\em Proceedings of the Designing Interactive Systems Conference},
  pages 126--135, 2012.

\bibitem{S13-RosenbloomMWM16}
R.~Rosenbloom, R.~A. Mason, H.~P. Wills, and B.~A. Mason.
\newblock Technology delivered self-monitoring application to promote
  successful inclusion of an elementary student with autism.
\newblock {\em Assistive Technology}, 28(1):9--16, 2016.

\bibitem{S14-MechlingAFB15}
L.~C. Mechling, K.~M. Ayres, A.~L. Foster, and K.~J. Bryant.
\newblock Evaluation of generalized performance across materials when using
  video technology by students with autism spectrum disorder and moderate
  intellectual disability.
\newblock {\em Focus on Autism and Other Developmental Disabilities},
  30(4):208--221, 2015.

\bibitem{S15-LiuSVS17}
Runpeng Liu, Joseph~P. Salisbury, Arshya Vahabzadeh, and Ned~T. Sahin.
\newblock {Feasibility of an Autism-Focused Augmented Reality Smartglasses
  System for Social communication and Behavioral coaching}.
\newblock {\em {FRONTIERS IN PEDIATRICS}}, {5}, {JUN 26} {2017}.

\bibitem{S16-TorradoGM17}
Juan~C. Torrado, Javier Gomez, and German Montoro.
\newblock {Emotional Self-Regulation of Individuals with Autism Spectrum
  Disorders: Smartwatches for Monitoring and Interaction}.
\newblock {\em {SENSORS}}, {17}({6}), {JUN} {2017}.

\bibitem{S17-CabiellesPPF17}
David Cabielles-Hernandez, Juan-Ramon Perez-Perez, MPuerto Paule-Ruiz, and
  Samuel Fernandez-Fernandez.
\newblock {Specialized Intervention Using Tablet Devices for Communication
  Deficits in Children with Autism Spectrum Disorders}.
\newblock {\em {IEEE TRANSACTIONS ON LEARNING TECHNOLOGIES}},
  {10}({2}):{182--193}, {APR-JUN} {2017}.

\bibitem{S18-ChevalierMIBT17}
Pauline Chevalier, Jean-Claude Martin, Brice Isableu, Christophe Bazile, and
  Adriana Tapus.
\newblock {Impact of sensory preferences of individuals with autism on the
  recognition of emotions expressed by two robots, an avatar, and a human}.
\newblock {\em {AUTONOMOUS ROBOTS}}, {41}({3, SI}):{613--635}, {MAR} {2017}.

\bibitem{S19-YingSA16}
Kuan~Tian Ying, Shahrul Badariah~Mat Sah, and Muhammad Haziq~Lim Abdullah.
\newblock {Personalised Avatar on Social Stories and Digital Storytelling:
  Fostering Positive Behavioural Skills for Children with Autism Spectrum
  Disorder}.
\newblock In {\em {2016 4TH INTERNATIONAL CONFERENCE ON USER SCIENCE AND
  ENGINEERING (I-USER)}}, {International Conference on User Science and
  Engineering}, pages {253--258}. {Univ Teknologi Mara; Int Islam Univ
  Malaysia; Univ Teknikal Malaysia Melaka; IEEE Malaysia Sect; IEEE Comp Soc;
  IEEE; IEEE Comp Soc Malaysia Sect}, {2016}.
\newblock {4th International Conference on User Science and Engineering
  (i-USEr), Melaka, MALAYSIA, AUG 23-25, 2016}.

\bibitem{S20-KhanPHHSM16}
Md. Nasfikur~R. Khan, Mohammad N.~H. Pias, K.~Habib, M.~Hossain, F.~Sarker, and
  K.~A. Mamun.
\newblock {Bolte Chai: An Augmentative and Alternative Communication Device for
  Enhancing Communication for Nonverbal Children}.
\newblock In {\em {2016 INTERNATIONAL CONFERENCE ON MEDICAL ENGINEERING, HEALTH
  INFORMATICS AND TECHNOLOGY (MEDITEC)}}. {United Int Univ; EMB Bangladesh
  Chapter; IEEE Bangladesh Sect; Medtronic; United Hospital; SIEMENS; IBN SINA
  TRUST; Tradevis Ltd; MARKS}, {2016}.
\newblock {1st International Conference on Medical Engineering, Health
  Informatics and Technology (MediTec), Dhaka, BANGLADESH, DEC 17-18, 2016}.

\bibitem{S21-SharmaSAVHKTR16}
Sumita Sharma, Saurabh Srivastava, Krishnaveni Achary, Blessin Varkey, Tomi
  Heimonen, Jaakko Hakulinen, Markku Turunen, and Nitendra Rajput.
\newblock {Promoting Joint Attention with Computer Supported Collaboration in
  Children with Autism}.
\newblock In {\em {ACM CONFERENCE ON COMPUTER-SUPPORTED COOPERATIVE WORK AND
  SOCIAL COMPUTING (CSCW 2016)}}, pages {1560--1571}. {Assoc Comp Machinery;
  ACM Special Interest Grp Human Interact}, {2016}.
\newblock {19th ACM Conference on Computer-Supported Cooperative Work and
  Social Computing (CSCW), San Francisco, CA, FEB 27-MAR 02, 2016}.

\bibitem{S22-MagriniSCC16}
Massimo Magrini, Ovidio Salvetti, Andrea Carboni, and Olivia Curzio.
\newblock {An Interactive Multimedia System for Treating Autism Spectrum
  Disorder}.
\newblock In {Hua, G and Jegou, H}, editor, {\em {COMPUTER VISION - ECCV 2016
  WORKSHOPS, PT II}}, volume {9914} of {\em {Lecture Notes in Computer
  Science}}, pages {331--342}, {2016}.
\newblock {14th European Conference on Computer Vision (ECCV), Amsterdam,
  NETHERLANDS, OCT 08-16, 2016}.

\bibitem{S23-HulusicP16}
Vedad Hulusic and Nirvana Pistoljevic.
\newblock {Read, Play and Learn: An Interactive E-book for Children with
  Autism}.
\newblock In {DeGloria, A and Veltkamp, R}, editor, {\em {Games and Learning
  Alliance, GALA 2015, Revised Selected Papers}}, volume {9599} of {\em
  {Lecture Notes in Computer Science}}, pages {255--265}. {Serious Games Soc;
  Univ Genoa}, {2016}.
\newblock {4th International Conference on Games and Learning (GALA), Sapienza
  Univ, Rome, ITALY, DEC 09-11, 2015}.

\bibitem{S24-SturmPP16}
Deborah Sturm, Ed~Peppe, and Bertram Ploog.
\newblock {eMot-iCan: Design of an Assessment Game for Emotion Recognition in
  Players with Autism}.
\newblock In {\em {2016 IEEE INTERNATIONAL CONFERENCE ON SERIOUS GAMES AND
  APPLICATIONS FOR HEALTH}}, {IEEE International Conference on Serious Games
  and Applications for Health}. {UCF; Univ Central Florida, Florida Interact
  Entertainment Acad; Inst Politecnico Cavado Ave, Escola Superior Tecnologia;
  IEEE; IEEE Comp Soc}, {2016}.
\newblock {IEEE International Conference on Serious Games and Applications for
  Health, Orlando, FL, MAY 11-13, 2016}.

\bibitem{S25-ChevalierIMT16}
Pauline Chevalier, Brice Isableu, Jean-Claude Martin, and Adriana Tapus.
\newblock {Individuals with Autism: Analysis of the First Interaction with Nao
  Robot Based on Their Proprioceptive and Kinematic Profiles}.
\newblock In {Borangiu, T}, editor, {\em {ADVANCES IN ROBOT DESIGN AND
  INTELLIGENT CONTROL}}, volume {371} of {\em {Advances in Intelligent Systems
  and Computing}}, pages {225--233}, {2016}.
\newblock {24th International Conference on Robotics in Alpe-Adria-Danube
  Region (RAAD), Bucharest, ROMANIA, MAY 27-29, 2015}.

\bibitem{S26-LandowskaS16}
Agnieszka Landowska and Michal Smiatacz.
\newblock {Mobile Activity Plan Applications for Behavioral Therapy of Autistic
  Children}.
\newblock In {Gruca, A and Brachman, A and Kozielski, S and Czachorski, T},
  editor, {\em {MAN-MACHINE INTERACTIONS 4, ICMMI 2015}}, volume {391} of {\em
  {Advances in Intelligent Systems and Computing}}, pages {115--125}. {IEEE
  Poland Sect Chapters; IEEE; Silesian Univ Technol, Inst Informat; Polish Acad
  Sci, Inst Theoret \& Appl Informat}, {2016}.
\newblock {4th International Conference on Man-Machine Interactions (ICMMI),
  Kocierz Pass, POLAND, OCT 06-09, 2015}.

\bibitem{S27-SkillenDNB16}
K.~L. Skillen, M.~P. Donnelly, C.~D. Nugent, and N.~Booth.
\newblock {LifePal: A Mobile Self-management Tool for Supporting Young People
  with Autism}.
\newblock In {Kyriacou, E and Christofides, S and Pattichis, CS}, editor, {\em
  {XIV MEDITERRANEAN CONFERENCE ON MEDICAL AND BIOLOGICAL ENGINEERING AND
  COMPUTING 2016}}, volume~{57} of {\em {IFMBE Proceedings}}, pages
  {1168--1173}. {Cyprus Assoc Med Phys \& Biomed Engn; Univ Cyprus; IFMBE;
  EAMBES; Frederick Univ; European Univ}, {2016}.
\newblock {14th Mediterranean Conference on Medical and Biological Engineering
  and Computing (MEDICON), Paphos, CYPRUS, MAR 31-APR 02, 2016}.

\bibitem{S28-KhoslaNC15}
Rajiv Khosla, Khanh Nguyen, and Mei-Tai Chu.
\newblock {Service Personalisation of Assistive Robot for Autism Care}.
\newblock In {\em {IECON 2015 - 41ST ANNUAL CONFERENCE OF THE IEEE INDUSTRIAL
  ELECTRONICS SOCIETY}}, {IEEE Industrial Electronics Society}, pages
  {2088--2093}. {IEEE Ind Elect Soc; SICE; Robot Soc Japan; IEEJ; IEEJ Cias;
  JSPE; JSME; EiC; SAE Japan}, {2015}.
\newblock {41st Annual Conference of the IEEE-Industrial-Electronics-Society
  (IECON), Yokohama, JAPAN, NOV 09-12, 2015}.

\bibitem{S29-MeiMQ15}
Chao Mei, Lee Mason, and John Quarles.
\newblock {How 3D Virtual Humans Built by Adolescents with ASD Affect Their 3D
  Interactions}.
\newblock In {\em {ASSETS'15: PROCEEDINGS OF THE 17TH INTERNATIONAL ACM
  SIGACCESS CONFERENCE ON COMPUTERS \& ACCESSIBILITY}}, pages {155--162}. {ACM
  SIGACCESS}, {2015}.
\newblock {17th International ACM SIGACCESS Conference on Computers and
  Accessibility (ASSETS 2015), Lisbon, PORTUGAL, OCT 26-28, 2015}.

\bibitem{S30-VoonBMJA15}
Nyuk~Hiong Voon, Siti~Nor Bazilah, Abdullah Maidin, Halina Jumaat, and
  Muhammad~Zulfadhli Ahmad.
\newblock {AutiSay: A Mobile Communication Tool for Autistic Individuals}.
\newblock In {PhonAmnuaisuk, S and Au, TW}, editor, {\em {COMPUTATIONAL
  INTELLIGENCE IN INFORMATION SYSTEMS}}, volume {331} of {\em {Advances in
  Intelligent Systems and Computing}}, pages {349--359}. {Bank Islam Brunei
  Darussalam; DST; Int Neural Network Soc}, {2015}.
\newblock {4th INNS Symposia on Computational Intelligencein Information
  Systems (INNS-CIIS), Inst Teknologi Brunei, Bandar Seri Begawan, BRUNEI, NOV
  07-09, 2014}.

\bibitem{S31-QidwaiSHM14}
Uvais Qidwai, Mohamed Shakir, Wa'ed Hakouz, and Nour Musa.
\newblock {Wirelessly Controlled Mimicing Humanoid Robot Assisting Children
  with ASD in learning Focused and Coordinated Behaviors}.
\newblock In {Bouras, A and Tari, Z and Erradi, A and Abdelwahed, S}, editor,
  {\em {2014 IEEE/ACS 11TH INTERNATIONAL CONFERENCE ON COMPUTER SYSTEMS AND
  APPLICATIONS (AICCSA)}}, {International Conference on Computer Systems and
  Applications}, pages {156--160}. {IEEE; ACS}, {2014}.
\newblock {11th IEEE/ACS International Conference on Computer Systems and
  Applications (AICCSA), Doha, QATAR, NOV 10-13, 2014}.

\bibitem{S32-GruarinWB13}
Alberto Gruarin, Michel~A. Westenberg, and Emilia~I. Barakova.
\newblock {StepByStep: Design of an Interactive Pictorial Activity Game for
  Teaching Generalization Skills to Children with Autism}.
\newblock In {Anacleto, JC and Clua, EWG and DaSilva, FSC and Fels, S and Yang,
  HS}, editor, {\em {ENTERTAINMENT COMPUTING - ICEC 2013}}, volume {8215} of
  {\em {Lecture Notes in Computer Science}}, pages {87--92}. {IFIP; Brazilian
  Ctr Support Res \& Educ; Sao Paulo Supporting Agcy Res; Graphics Animat \&
  New Media Ctr Canada}, {2013}.
\newblock {12th IFIP International Conference on Entertainment Computing
  (ICEC), Sao Paulo, BRAZIL, OCT 16-18, 2013}.

\bibitem{S33-DimitrovaVB12}
Maya Dimitrova, Niko Vegt, and Emilia Barakova.
\newblock {Designing a System of Interactive Robots for Training Collaborative
  Skills to Autistic Children}.
\newblock In {\em {2012 15TH INTERNATIONAL CONFERENCE ON INTERACTIVE
  COLLABORATIVE LEARNING (ICL)}}, {2012}.
\newblock {15th International Conference on Interactive Collaborative Learning
  (ICL), Villach, AUSTRIA, SEP 26-28, 2012}.

\bibitem{S1-IWVR/ParesCDFFGS04}
N~Par{\'{e}}s, A~Carreras, J~Durany, J.~Ferrer, P.~Freixa, D.~G{\'{o}}mez, and
  A.~Sanjurjo.
\newblock Mediate: An interactive multisensory environment for children with
  severe autism and no verbal communication.
\newblock In {\em Proc. of the Third International Workshop on Virtual
  Rehabilitation}, 2004.

\end{thebibliography}


\begin{thebibliography}{10}

\bibitem{APA}
{American Psychiatric Association}.
\newblock Diagnostic and statistical manual of mental disorders (5th ed.),
  2013.

\bibitem{Baird2006}
G.~Baird, E.~Simonoff, A.~Pickles, S.~Chandler, T.~Loucas, D.~Meldrum, and
  T.~Charman.
\newblock {Prevalence of disorders of the autism spectrum in a population
  cohort of children in South Thames: The special needs and autism project
  (SNAP)}.
\newblock {\em Lancet}, 68(9531):210--215., 2006.

\bibitem{DBLP:journals/tse/BatorySR04}
D.~S. Batory, J.~N. Sarvela, and A.~Rauschmayer.
\newblock Scaling step-wise refinement.
\newblock {\em IEEE Trans. Software Eng.}, 30(6):355--371, 2004.

\bibitem{Cohen2006}
H.~Cohen, M.~Amerine-Dickens, and T.~Smith.
\newblock Early intensive behavioral treatment: Replication of the ucla model
  in a community setting.
\newblock {\em Developmental and Behavioral Pediatrics}, 27(2):145–155, 2006.

\bibitem{ANYAS/ElKalioubyPB06}
R.~el~Kaliouby, R.~Picard, and S.~Baron-Cohen.
\newblock {Affective computing and autism}.
\newblock {\em Annals of the New York Academy of Sciences}, 1093(1):228--248,
  2006.

\bibitem{SAGE/GrynszpanWPG14}
O.~Grynszpan, P.~L.~T. Weiss, F.~Perez-Diaz, and E.~Gal.
\newblock Innovative technology-based interventions for autism spectrum
  disorders: A meta-analysis.
\newblock {\em Autism}, 18(4):346--361, 2014.
\newblock PMID: 24092843.

\bibitem{Kientz2013}
J.~A. Kientz, M.~S. Goodwin, G.~R. Hayes, and G.~D. Abowd.
\newblock {\em {Interactive Technologies for Autism}}.
\newblock Morgan {\&} Claypool Publishers, 2013.

\bibitem{Lorah2014}
E.~R. Lorah, A.~Parnell, P.~S. Whitby, and D.~Hantula.
\newblock A systematic review of tablet computers and portable media players as
  speech generating devices for individuals with autism spectrum disorder.
\newblock {\em Journal of Autism and Developmental Disorders}, pages 1--13,
  2014.

\bibitem{Odom2010}
S.~L. Odom, B.~A. Boyd, L.~J. Hall, and K.~Hume.
\newblock Evaluation of comprehensive treatment models for individuals with
  autism spectrum disorders.
\newblock {\em Journal of Autism and Developmental Disorders}, 40(4):425--436,
  2010.

\bibitem{SPLE}
K.~Pohl, G.~Bockle, and F.~J. van~der Linden.
\newblock {\em Software {P}roduct {L}ine {E}ngineering: {F}oundations,
  {P}rinciples and {T}echniques}.
\newblock Springer, 2005.

\bibitem{Rogers2010}
S.~J. Rogers and G.~Dawson.
\newblock {\em Early start Denver model for young children with autism:
  Promoting language, learning, and engagement.}
\newblock Guilford Press, 2010.

\bibitem{Stephenson2015}
J.~Stephenson and L.~Limbrick.
\newblock A review of the use of touch-screen mobile devices by people with
  developmental disabilities.
\newblock {\em Journal of Autism and Developmental Disorders},
  45(12):3777--3791, 2015.

\bibitem{WohlinExperimentalSE12}
C.~Wohlin, P.~Runeson, M.~H{\"o}st, M.~C. Ohlsson, and B.~Regnell.
\newblock {\em Experimentation in Software Engineering}.
\newblock Springer, 2012.

\bibitem{Wong2015}
C.~Wong, S.~L. Odom, K.~A. Hume, A.~W. Cox, A.~Fettig, S.~Kucharczyk, M.~E.
  Brock, J.~B. Plavnick, V.~P. Fleury, and T.~R. Schultz.
\newblock {Evidence-Based Practices for Children, Youth, and Young Adults with
  Autism Spectrum Disorder: A Comprehensive Review}.
\newblock {\em Journal of Autism and Developmental Disorders},
  45(7):1951--1966, 2015.

\end{thebibliography}


\bibliographystyleS{unsrt}
\bibliographyS{primary-sources}

\end{document}